\documentclass{article}
\usepackage{arxiv}
\usepackage{fontspec}
\usepackage{amsmath,unicode-math}
\newcommand{\matmul}{\mathbin{@}}
\usepackage{booktabs,longtable,array}
\usepackage{graphicx}
\usepackage{calc}
\usepackage{xcolor,listings}
\definecolor{CodeBackground}{HTML}{F5F6F8}
\definecolor{CodeRule}{HTML}{B8C4D0}
\definecolor{CodeKeyword}{HTML}{264B70}
\definecolor{CodeString}{HTML}{70513C}
\lstdefinestyle{paper}{
  basicstyle=\ttfamily\fontsize{9}{10.5}\selectfont,
  columns=fullflexible,keepspaces=true,showstringspaces=false,
  upquote=true,breaklines=true,breakatwhitespace=false,
  backgroundcolor=\color{CodeBackground},
  frame=l,rulecolor=\color{CodeRule},framerule=0.6pt,framesep=4pt,
  xleftmargin=6pt,xrightmargin=4pt,
  aboveskip=4pt,belowskip=6pt,
  keywordstyle=\color{CodeKeyword}\bfseries,
  stringstyle=\color{CodeString},commentstyle=\color{gray}\itshape,
  tabsize=4
}
\usepackage[font=small,labelfont=bf]{caption}
\usepackage{xurl}
\usepackage[hidelinks,unicode]{hyperref}
\hypersetup{pdftitle={Algebraic Retrieval: Composable Search for Agents},pdfauthor={Damian Delmas}}
\providecommand{\tightlist}{\setlength{\itemsep}{0pt}\setlength{\parskip}{0pt}}
\renewcommand{\shorttitle}{Algebraic Retrieval}
\title{Algebraic Retrieval: Composable Search for Agents}
\author{Damian Delmas \\
Independent Researcher, Vancouver, BC \\
\texttt{damian@getflex.dev}}
\date{}
\begin{document}
\maketitle
\begin{abstract}
Algebraic Retrieval lets AI agents compose search strategies at query time. Relevance criteria, eligibility constraints, and ranking preferences can be expressed together in a mathematical query. The query surface exposes available operations, so an agent can combine them for the question at hand and revise a program after inspecting results. We evaluate execution parity, not agent behavior or retrieval quality. Building on Programmatic Embedding Modulation (PEM), which exposes vector and score arithmetic during retrieval, we demonstrate contrastive scoring, candidate-pool reranking, and weighted ranking as composable queries, alongside executable SQL and PyTerrier counterparts. On the public 11,429-document Vaswani fixture, each program's implementations select the same document set with score differences below \texttt{1e-6}; one tied pair orders differently across scoring paths.
\end{abstract}

\hypertarget{introduction}{%
\section{Introduction}\label{introduction}}

Retrieval is often presented to an agent as one search call with a query and a few parameters. An agent seeking implementation architecture may also want to suppress marketing material, restrict candidates to engineering discussions, and weight scores by recency. These choices can be written together as one expression.

Algebraic Retrieval lets the agent compose that expression at query time. It chooses scoring directions, eligible records, and per-record weights, then can revise the expression after inspecting results.

Programmatic Embedding Modulation (PEM) {[}5{]} exposes vectors and scores for arithmetic at query time. Algebraic Retrieval gives that surface a mathematical query notation. We compare three programs with SQL and PyTerrier {[}3{]}, a Python library for composing retrieval pipelines.

\hypertarget{preliminaries}{%
\section{Preliminaries}\label{preliminaries}}

An embedding represents a record or query as a vector of numbers in a shared space. Let \(E\) hold one record vector per row and \(q\) be a query vector. Matrix multiplication \(E\matmul q\) produces one score per record. With unit-length record and query vectors, these scores equal cosine similarity. The scores and their record identities form a scored relation \(S\).

PEM {[}5{]} exposes the embedding matrix and score array for arithmetic before selection. A caller can change the query vector before scoring, or add, subtract, and weight scores afterward. SQL filters choose the participating records. Algebraic Retrieval gives these operations a mathematical query notation. Its contract uses normalized stored and primitive query vectors but leaves composed queries unnormalized.

For query vectors in the same embedding space, \(+\) and \(-\) combine their scores, and scalar \texttt{*} scales a contribution. For example, \((E\matmul q_1)-0.5\,(E\matmul q_2)\) favors the first direction while suppressing the second. Selection happens after composition; subtracting two already-truncated top-ten lists would be a different operation. In displayed formulas, \(q_1\) and \(q_2\) denote the source tokens \texttt{q1} and \texttt{q2}.

A mask \(m\) is a set of eligible identities. \(m\triangleright S\) removes other records from a scored relation \(S\). A weight relation \(w\) supplies a scalar for each eligible identity, and \(w\odot S\) multiplies its score. Here \(w\) covers every eligible identity; a missing weight is represented as zero. Masks remove records; a zero weight leaves a record present with score zero. Finally, \(\tau_{10}(S)\) selects up to ten results, ordered by score descending and identity ascending. The ASCII spelling is \texttt{top(10,\ S)}.

\hypertarget{architecture}{%
\section{Architecture}\label{architecture}}

The runtime uses a materializer: code that executes retrieval operations and stores the results in a temporary SQL table. It resolves named operands through SQLite, while NumPy performs vector arithmetic. The parser reads the expression; the planner chooses its execution steps.

\begin{figure}[!ht]
\centering
\includegraphics[width=\linewidth]{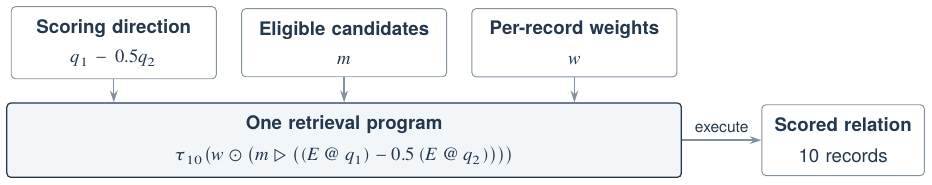}
\caption{An agent combines scoring directions, a candidate mask, and per-record weights into the query of §5.3. Execution returns a scored relation that can be inspected or composed further.}
\label{fig:retrieval-program}
\end{figure}

The agent submits a program through \texttt{algebra('<expression>')}, which can also appear inside a larger SQL query to join results to record text or metadata. Changing a coefficient or adding a mask changes the retrieval strategy through the same call.

Because the query vectors remain explicit in the expression, the planner can combine them before scoring. It rewrites two scoring passes as one using:

\[
  (E\matmul q_1)-0.5\,(E\matmul q_2)
  = E\matmul(q_1-0.5q_2).
\]

The rewrite preserves the mathematical (unnormalized) scores and uses one matrix multiplication. It is exact in real arithmetic; 32-bit floating-point (\texttt{float32}) rounding can cause small differences.

For a fixed \(E\), changing \(q\) leaves the scored identities intact. A mask changes which records participate, while per-record weighting generally cannot be absorbed into a change of \(q\).

The SQL listings are executable equivalents, not compiler output. They calculate cosine similarity with the SQLite extension sqlite-vec {[}10{]}, binding \texttt{:q1} and \texttt{:q2} to the query vectors. They query \texttt{embeddings(docno, vector)} and the views \texttt{m(id)} and \texttt{w(id, weight)} defined in Example bindings.

\hypertarget{orient-what-the-agent-sees}{%
\section{Orient: what the agent sees}\label{orient-what-the-agent-sees}}

Before writing a query, the agent calls:

\begin{lstlisting}
algebra('orient')
\end{lstlisting}

The response lists available operands, their source relations, and whether they are usable. For this comparison, the proof runner supplies the following fixture-specific bindings; a live contract-backed cell returns the same kind of rows:

\begin{longtable}[]{@{}llll@{}}
\toprule\noalign{}
section & token & domain & status \\
\midrule\noalign{}
\endhead
\bottomrule\noalign{}
\endlastfoot
relation & \texttt{R} & \texttt{documents(docno,\ text)} & \texttt{ok} \\
matrix & \texttt{E} & \texttt{embeddings(docno,\ vector)} & \texttt{ok} \\
query & \texttt{q1} & \texttt{query\_vectors(qid,\ vector)} & \texttt{ok} \\
query & \texttt{q2} & \texttt{embeddings(docno,\ vector)} & \texttt{ok} \\
mask & \texttt{m} & \texttt{pyterrier\_bm25\_results(docno,\ qid)} & \texttt{ok} \\
weight & \texttt{w} & \texttt{pyterrier\_bm25\_results(docno,\ score,\ qid)} & \texttt{ok} \\
\end{longtable}

The live response also advertises the operators used below. Whoever integrates the corpus declares how each symbol maps to data; the example runner declares the bindings used here. The agent composes expressions over those named inputs.

The examples use the 11,429-document \texttt{ir\_datasets:vaswani} collection {[}4{]}, with deterministic 128-dimensional signed-hash vectors and normalized stored rows. \texttt{q1} is the stored query vector for “What are chemical reactions?”; \texttt{q2} is a fixed vector from document \texttt{2008}. BM25 ranks documents by query-term matches; the mask contains the 52 candidates in its recorded PyTerrier result for that query. Each weight is its BM25 score divided by the maximum score in that result. The vectors make the computation reproducible; they are not a learned-embedding retrieval evaluation.

\newpage

\hypertarget{queries-the-agent-can-compose}{%
\section{Queries the agent can compose}\label{queries-the-agent-can-compose}}

The examples introduce contrastive (signed) scoring and candidate restriction, then combine both with per-record weights. In PyTerrier, \texttt{>>} passes results to the next stage and \texttt{\% 10} keeps the top ten. \texttt{D} supplies every document, \texttt{M} supplies the recorded BM25 candidates, \texttt{A} and \texttt{B} score with the two query vectors, and \texttt{W} applies the declared weights. \texttt{T} orders by score and document identity before cutoff. Their definitions appear at the end of the paper. Reported score differences are relative to Algebra.

\hypertarget{subtract-one-similarity-search-from-another}{%
\subsection{Subtract one similarity search from another}\label{subtract-one-similarity-search-from-another}}

The agent can use a second query or an example document to specify an unwanted direction. Here document \texttt{2008} fixes the subtraction operand reproducibly; it is not assigned an irrelevance judgment.

\textbf{Algebraic expression}

\[
  \tau_{10}\!\left((E\matmul q_1)-0.5\,(E\matmul q_2)\right)
\]

\textbf{Equivalent SQL}

\begin{lstlisting}[language=SQL]
SELECT docno AS id,
       (1 - vec_distance_cosine(vector, :q1))
       - 0.5 * (1 - vec_distance_cosine(vector, :q2)) AS score
FROM embeddings
ORDER BY score DESC, id ASC
LIMIT 10;
\end{lstlisting}

\textbf{PyTerrier}

\begin{lstlisting}[language=Python]
(D >> (A + (-0.5) * B) >> T) % 10
\end{lstlisting}

PyTerrier adds the first scorer to a negatively weighted second scorer. The SQL alternative calculates two cosine similarities per row; Algebra combines the vectors and scores once. All select the same ten-document set, with document \texttt{9448} first. PyTerrier returns the same order, with a maximum score difference of \texttt{2.24e-8}. Documents \texttt{7429} and \texttt{9217} tie exactly in Algebra and PyTerrier, where identity ordering places \texttt{7429} first. The SQL path separates them by approximately \texttt{5.96e-8} and places \texttt{9217} first; its maximum returned-score difference is also approximately \texttt{5.96e-8}.

Replacing subtraction with addition blends the two scoring directions.

\hypertarget{rerank-a-candidate-pool}{%
\subsection{Rerank a candidate pool}\label{rerank-a-candidate-pool}}

The agent restricts vector scoring to an existing first-pass result. Here the mask is the 52-document BM25 candidate pool exposed by \texttt{orient}.

\textbf{Algebraic expression}

\[
  \tau_{10}\!\left(m\triangleright(E\matmul q_1)\right)
\]

\textbf{Equivalent SQL}

\begin{lstlisting}[language=SQL]
SELECT e.docno AS id, 1 - vec_distance_cosine(e.vector, :q1) AS score
FROM embeddings AS e JOIN m ON m.id = e.docno
ORDER BY score DESC, id ASC
LIMIT 10;
\end{lstlisting}

\textbf{PyTerrier}

\begin{lstlisting}[language=Python]
(M >> A >> T) % 10
\end{lstlisting}

The candidate pool is scored again and then reduced to ten results. The Algebra planner applies the mask before its vector scoring pass. The SQL join expresses the same allowed identities, and PyTerrier passes those candidates to the scorer. All return the same ordered identities, with document \texttt{2417} first. PyTerrier scores match Algebra exactly; the SQL scores have a maximum observed difference of approximately \texttt{2.98e-8}.

\newpage

\hypertarget{compose-the-full-retrieval-strategy}{%
\subsection{Compose the full retrieval strategy}\label{compose-the-full-retrieval-strategy}}

Within the candidate pool, the agent combines target and suppression scores, applies the declared weights, and selects the top ten. Normalized BM25 weights serve as a mechanical demonstration.

\textbf{Algebraic expression}

\[
  \tau_{10}\!\left(
    w\odot\left(m\triangleright
      \left((E\matmul q_1)-0.5\,(E\matmul q_2)\right)
    \right)
  \right)
\]

\textbf{Equivalent SQL}

\begin{lstlisting}[language=SQL]
SELECT e.docno AS id,
       w.weight * ((1 - vec_distance_cosine(e.vector, :q1))
                   - 0.5 * (1 - vec_distance_cosine(e.vector, :q2))) AS score
FROM embeddings AS e
JOIN m ON m.id = e.docno JOIN w ON w.id = e.docno
ORDER BY score DESC, id ASC
LIMIT 10;
\end{lstlisting}

\textbf{PyTerrier}

\begin{lstlisting}[language=Python]
(M >> (A + (-0.5) * B) >> W >> T) % 10
\end{lstlisting}

All 52 eligible identities have weights. All three forms return the same ordered identities, with document \texttt{10703} first. Maximum score differences relative to Algebra are approximately \texttt{2.98e-8} for SQL and \texttt{1.49e-8} for PyTerrier.

The third program composes the choices introduced earlier: combine scoring directions, restrict the candidates, and apply per-record preferences before selection. Its output remains a scored relation that the agent can inspect and reuse.

\hypertarget{explanation-and-evidence}{%
\section{Explanation and evidence}\label{explanation-and-evidence}}

The composition works because intermediate scores remain paired with identities. Arithmetic changes scores, masks determine which records participate, and top-k selects the final result.

The PyTerrier 1.1.2 programs use scores independently computed by FAISS 1.15.0 \texttt{IndexFlatIP}, an exact inner-product scorer. The SQL alternatives use sqlite-vec 0.1.9. All document sets match Algebra and score differences are below \texttt{1e-6}; §5.1 reports the ordering difference. The masked cases use the recorded upstream BM25 pool; the unmasked contrastive case scores all 11,429 rows. These checks do not compare timing or BM25 implementations.

Score tolerance does not imply rank agreement. Algebra and PyTerrier tie \texttt{7429} and \texttt{9217} at \texttt{0.24868088960647583}, a reference gap of zero; SQL favors \texttt{9217} by approximately \texttt{5.96e-8}. Identity ordering resolves only computed ties. If each alternative score differs from its reference by at most \(\varepsilon\), a reference gap greater than \(2\varepsilon\) preserves their order.

SQL deviations of approximately \texttt{2.98e-8} also occur in §5.2, which uses a single query vector. A two-pass NumPy calculation of §5.1 preserves the tie, so the ordering difference does not require the rewrite. Stored float32 rows are approximately unit length; sqlite-vec 0.1.9 {[}10{]} recomputes vector lengths using float32 accumulation, then rounds cosine distance before SQL subtracts it from one. Reproducing those steps reproduces the pair's split: a deterministic rounding effect.

The repository's \href{https://github.com/algebraicretrieval/algebraicretrieval/tree/main/proofs/pyterrier}{six-case validation} also checks score cutoffs, which retain records that meet a chosen minimum score. The \href{https://github.com/algebraicretrieval/algebraicretrieval/blob/main/tests/test_contract.py}{portable compiler witness} exercises support, signed scores, weights, and cutoff ties; companion tests check \href{https://github.com/algebraicretrieval/algebraicretrieval/blob/main/tests/test_typed_operators.py}{restriction versus zero-weighting on negative scores} and \href{https://github.com/algebraicretrieval/algebraicretrieval/blob/main/tests/test_score_semantics.py}{tie ordering, the rewrite in §3, and score cutoffs}. A \href{https://github.com/algebraicretrieval/algebraicretrieval/blob/main/tests/test_barrier_semantics.py}{feedback test} selects the highest-scoring record, uses its vector as a new query, and scores again. Applying the mask before that selection changes the final ordering, so the planner must preserve where restriction occurs.

PyTerrier composes retrieval stages, called transformers {[}3{]}, and supports \href{https://pyterrier.readthedocs.io/en/latest/transformer.html\#optimisation}{pipeline optimizations}. Here the agent writes one expression over discovered operands; the matrix and query vectors remain explicit for rewrites such as §3's. Earlier work develops vector scoring and query revision from example documents {[}1,2{]}, and top-k selection over database relations {[}6,7{]}. Vespa lets users define document-scoring expressions {[}8{]}; LOTUS uses language models to evaluate natural-language conditions on table rows {[}9{]}. The experiments do not measure retrieval-quality gains, approximate nearest-neighbor recall, comparative agent query-writing accuracy, or universal compiler correctness.

\hypertarget{conclusion}{%
\section{Conclusion}\label{conclusion}}

An agent can write retrieval as arithmetic over similarity scores, restrict it with a mask, and change the program at query time. The existing runtime executes these short expressions and returns relations the agent can continue to compose. On the public fixture, three implementations select the same document sets with score differences below \texttt{1e-6}, yet one tied pair orders differently. Numerical agreement must be checked separately from rank agreement.

\newpage

\hypertarget{references}{%
\section*{References}\label{references}}
\addcontentsline{toc}{section}{References}

\begin{enumerate}
\raggedright
\def\labelenumi{\arabic{enumi}.}
\tightlist
\item
  G. Salton, A. Wong, and C. S. Yang. ``A Vector Space Model for Automatic Indexing.'' \emph{Communications of the ACM}, 1975.
\item
  J. J. Rocchio. ``Relevance Feedback in Information Retrieval.'' \emph{The SMART Retrieval System}, 1971.
\item
  Craig Macdonald and Nicola Tonellotto. ``Declarative Experimentation in Information Retrieval using PyTerrier.'' \emph{ICTIR}, 2020.
\item
  Sean MacAvaney et al.~``Simplified Data Wrangling with ir\_datasets.'' \emph{SIGIR}, 2021.
\item
  Damian Delmas. ``flexvec: SQL Vector Retrieval with Programmatic Embedding Modulation.'' arXiv:2603.22587, 2026. DOI: 10.48550/arXiv.2603.22587.
\item
  Chengkai Li, Kevin C.-C. Chang, Ihab F. Ilyas, and Sumin Song. ``RankSQL: Query Algebra and Optimization for Relational Top-k Queries.'' \emph{SIGMOD}, 2005. DOI: 10.1145/1066157.1066173.
\item
  Ronald Fagin, Amnon Lotem, and Moni Naor. ``Optimal Aggregation Algorithms for Middleware.'' \emph{PODS}, 2001. DOI: 10.1145/375551.375567.
\item
  Vespa. ``YQL Query Language,'' ``Ranking Expressions,'' and ``Phased Ranking.'' Vespa Documentation, accessed 2026.
\item
  Liana Patel, Siddharth Jha, Melissa Pan, Harshit Gupta, Parth Asawa, Carlos Guestrin, and Matei Zaharia. ``Semantic Operators and Their Optimization: Enabling LLM-Based Data Processing with Accuracy Guarantees in LOTUS.'' \emph{PVLDB}, 2025.
\item
  Alex Garcia. ``sqlite-vec: API Reference.'' https://alexgarcia.xyz/sqlite-vec/api-reference.html. Version 0.1.9 used in the SQL comparisons; \href{https://github.com/asg017/sqlite-vec/blob/v0.1.9/sqlite-vec.c\#L436}{cosine implementation}.
\end{enumerate}

\hypertarget{example-bindings}{%
\section*{Example bindings}\label{example-bindings}}
\addcontentsline{toc}{section}{Example bindings}

The bindings use PyTerrier (\texttt{pt}), pandas (\texttt{pd}), and NumPy (\texttt{np}). \texttt{ids} lists every document; \texttt{score1} and \texttt{score2} map identities to their FAISS scores for the two query vectors. \texttt{bm25} contains the recorded first-pass result. \texttt{topics} contains the single fixture query, passed to each pipeline through \texttt{.transform(topics)}.

\begin{lstlisting}[language=Python]
weight = dict(zip(bm25.docno, bm25.score / bm25.score.max()))
D = pt.Transformer.from_df(pd.DataFrame({"qid": "1", "docno": ids}))
M = pt.Transformer.from_df(bm25)
A = pt.apply.doc_score(lambda r: float(score1[r["docno"]]))
B = pt.apply.doc_score(lambda r: float(score2[r["docno"]]))
W = pt.apply.doc_score(lambda r: r["score"] * weight[r["docno"]])

def score_id_order(frame):
    out = frame.sort_values(
        ["score", "docno"], ascending=[False, True], kind="mergesort"
    ).reset_index(drop=True)
    out["rank"] = np.arange(len(out))
    return out

T = pt.apply.generic(score_id_order)
\end{lstlisting}

The explicit ordering stage supplies the comparison's tie policy. PyTerrier's default tied ranks follow input order. Code, fixture instructions {[}4{]}, and dependencies are available at \url{https://github.com/algebraicretrieval/algebraicretrieval/tree/main/proofs/pyterrier}. After building the fixture and installing the pinned dependencies described there, run the \href{https://github.com/algebraicretrieval/algebraicretrieval/blob/main/proofs/pyterrier/examples.py}{\texttt{examples.py}} comparison from the repository root:

\begin{lstlisting}[language=bash]
python proofs/pyterrier/examples.py
\end{lstlisting}

The native SQL alternatives bind \texttt{:q1} to the stored vector for query \texttt{1} and \texttt{:q2} to document \texttt{2008}'s embedding, and use these views. SQL's window maximum and PyTerrier's \texttt{bm25.score.max()} both cover the same 52 rows with \texttt{qid=\textquotesingle{}1\textquotesingle{}}; the maximum is \texttt{22.076426496954774}, and the normalized weights agree exactly.

\begin{lstlisting}[language=SQL]
CREATE TEMP VIEW m AS
SELECT docno AS id FROM pyterrier_bm25_results WHERE qid='1';
CREATE TEMP VIEW w AS
SELECT docno AS id, score / max(score) OVER () AS weight
FROM pyterrier_bm25_results WHERE qid='1';
\end{lstlisting}

\end{document}